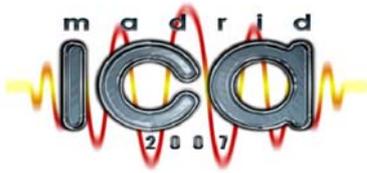

# NOISE REDUCTION COMBINING MICROPHONE AND PIEZOELECTRIC DEVICE

Takahashi, Naoya[1]; Matsumoto, Mitsuharu[2]; Hashimoto, Shuji[3]
[1]Waseda University; 3-4-1 Okubo, Shinjuku-ku, Tokyo, Japan; takanao@moegi.waseda.jp
[2]Waseda University; 3-4-1 Okubo, Shinjuku-ku, Tokyo, Japan; matsu@shalab.phys.waseda.ac.jp
[3]Waseda University; 3-4-1 Okubo, Shinjuku-ku, Tokyo, Japan; shuji@waseda.jp

**ABSTRACT**
It is often required to extract the sound of an objective instrument played in concert with other instruments. Microphone array is one of the effective ways to enhance a sound from a specific direction. However it is not effective in an echoic room such as concert hall. The pickup microphone attached on the specific musical instrument is often employed to obtain the sound exclusively from other instrumental sounds. The obtained timbre differ from the one we hear at the usual listening position. The purpose of this paper is to propose a new method of sound separation that utilizes the piezoelectric device attached on the body of the instrument. The signal from the attached device has a different spectrum from the sound heard by the audience but has the same frequency components as the instrumental sound. Our idea is to use the device signal as a modifier of the sound focusing filter applied to the microphone sound at the listening position. The proposed method firstly estimates the frequency components of the signal from the piezoelectric device. The frequency characteristics for filtering the microphone sound are changed so that it pass the estimated frequency components. Thus we can extract the target sound without distortion. The proposed method is a sort of dynamic sparseness approach. It was found that *SNR* is improved by 8.7dB through the experiments.

## 1. INTRODUCTION
In these days, the style of concert becomes diverse. When we play the instruments whose sound volumes are different, we need to amplify the sound volumes of soft sound volume instruments. In this case, it is sometimes required to extract the sound of a specific instrument played in concert with other instruments. Delay-sum type of microphone array [1] [2] [3] [4] is one of the effective ways to extract a sound from a specific direction. However, it needs lots of microphones to reduce the noise effectively. Adaptive microphone array [5] [6] is another method to reduce the noise. However, it is not effective to extract a sound in an echoic room such as concert hall. Moreover, the conventional microphone array methods can not also be applied in case that the noise and the signal come from the same direction. ICA [7] [8] is the method for the sound separation utilizing the independence of sound sources. However, It can not be applied to the nonstationary sound. The binary mask [9] [10] [11] is also effective ways to extract a sound. The binary mask requires two microphones at different positions to obtain different power ratio of the signal and the noise. However it is often difficult to set the microphone at an appropriate position for binary mask in concert hall. Binary mask is not applicable in case that the frequency components of the signal and the noise overlap.
In this paper, we propose noise reduction combining a microphone and a piezoelectric pickup. The piezoelectric pickup does not obtain the background sound. The proposed method firstly estimates the frequency components of the signal utilizing the signal from the piezoelectric device. Then, we filter the microphone sound to pass the estimated frequency components.
We explain the proposal method in Sec. 2. Some experimental results are given in Sec. 3.

## 2. PROPOSAL METHOD
### 2.1 Noise reduction combining microphone and piezoelectric pickup
In the proposal method, we utilize a pair of the microphone and a piezoelectric pickup. A piezoelectric device is made of ceramic. When a piezoelectric device is distorted, the voltage is generated on the surface. By utilizing this property, the piezoelectric device is utilized as a pickup which senses the vibration of the instrumental body. The signal from the piezoelectric pickup has a different spectrum from the sound heard by the audience but has the same frequency components as the instrumental sound because the pickup detects the vibration of the body that is the source of the instrumental sound.

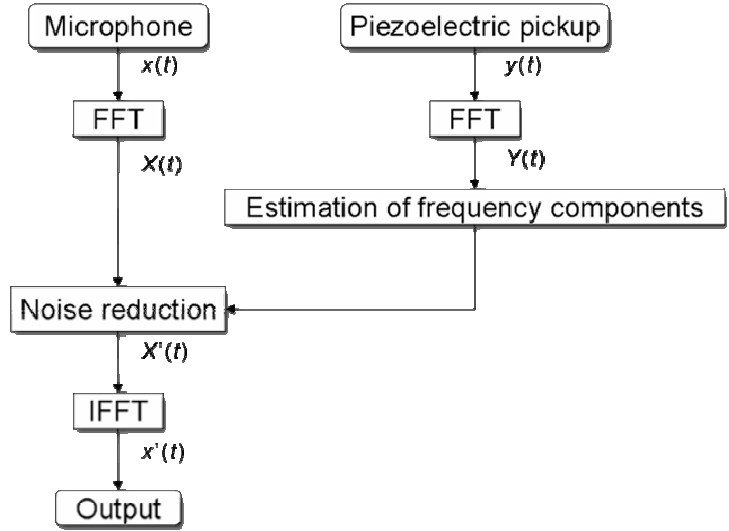

Figure 1. Flow chart of proposal method

Fig.1 shows the flow chart of the proposal method. We extract the objective instrumental sound utilizing the microphone and the piezoelectric pickup. Let us define $x(t)$ and $y(t)$ as the signal from the microphone and the piezoelectric device at time $t$, respectively. Then $x(t)$ and $y(t)$ are represented as follows:

$$x(t) = m_1(t) * s(t) + \sum_i m_{2i}(t) * n_i(t) \quad \text{(Eq.1)}$$

$$y(t) = p_1(t) * s(t) + \sum_i p_{2i}(t) * n_i(t) \quad \text{(Eq.2)}$$

where $*$ represents the convolution. $s(t)$ represents the signal, that is, the objective instrument sounds. $n_i(t)$ represents the $i$th noise, that is, other back ground sounds. $m_1(t)$ and $m_{2i}(t)$ represent the impulse responses from the signal sound source and the $i$th noise sound source to the microphone, respectively. $p_1(t)$ and $p_{2i}(t)$ represent the impulse responses from the signal sound source and the $i$th noise sound source to the piezoelectric pickup, respectively. We first transform the input signal $x(t)$ and $y(t)$ to the complex spectra $X(\tau, \omega)$ and $Y(\tau, \omega)$ as functions of the time frame $\tau$ and angular frequency $\omega$ by short time Fourier transformation(STFT) as follows:

$$X(\tau,\omega) = M_1(\omega)S(\tau,\omega) + \sum_i M_{2i}(\omega)N_i(\tau,\omega) \quad \text{(Eq.3)}$$

$$Y(\tau,\omega) = P_1(\omega)S(\tau,\omega) + \sum_i P_{2i}(\omega)N_i(\tau,\omega) \quad \text{(Eq.4)}$$

where $S(\tau, \omega)$ and $N(\tau, \omega)$ represent the complex spectra of $s(t)$ and $n_i(t)$, respectively. $M_1(\omega)$ and $M_{2i}(\omega)$ represent the transfer function of the microphone regarding the signal and the noise, respectively. $P_1(\omega)$ and $P_{2i}(\omega)$ represent the transfer function of the piezoelectric pickup regarding the signal and the noise, respectively. Our aim is to extract $m_1(t) * s(t)$ from $x(t)$.

We firstly explain the disjoint case. We assume that the frequency components of the signal and noise are disjoint. This assumption is described as follows:

$$S(\tau, \omega)N_i(\tau, \omega) = 0 \quad (\forall \; \omega, i) \quad \text{(Eq. 5)}$$

If the objective instrument does not resonate with other instruments sounds, the noise contained in $y(t)$ can be assumed as zero. This assumption can be described as follows:

$$P_{2i}(\omega)N_i(\tau, \omega) = 0 \quad (\forall \; \omega, i) \quad \text{(Eq. 6)}$$

Therefore, we can estimate the frequency components of the target signal utilizing the piezoelectric pickup signal by gating the power of frequency components as follows:

$$X'(\tau,\omega) = \begin{cases} X(\tau,\omega) & Y(\tau,\omega) \neq 0 \\ 0 & Y(\tau,\omega) = 0 \end{cases} \quad \text{(Eq. 7)}$$

$$= M_1(\omega)S(\tau,\omega)$$

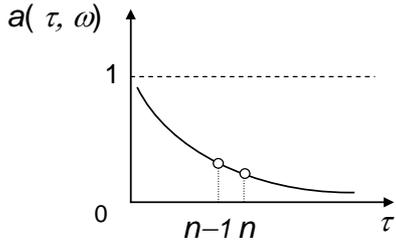 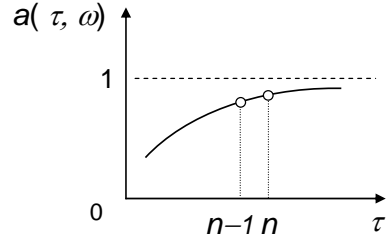

Figure 2. Updating of $a(\tau,\omega)$ in case of noise   Figure 3. Updating of $a(\tau,\omega)$ in case of the signal

where $X'(\tau, \omega)$ represents the processed signal of the proposal method. Finally, $X'(\tau, \omega)$ is transformed into $x'(t)$ by Inverse Fast Fourier Transformation (IFFT).
In reality, although the piezoelectric pickup may contain some noises, the power of the noise is sufficiently small. Hence, we can assume that the signal and the noise satisfy the following condition.

$$| P_{2i}(\omega) N_i(\tau, \omega) | < th \qquad (\forall \omega, i) \qquad \text{(Eq. 8)}$$
$$\exists \omega \ | P_1(\omega) S(\tau, \omega) | \neq 0 \Rightarrow | S'(\tau, \omega) | > th \qquad \text{(Eq. 9)}$$

where $th$ represents the threshold value. Eq. 8 means that the power of the noise contained in the signal from piezoelectric pickup is smaller than $th$. Eq. 9 means that the power of the signal contained in the signal from the piezoelectric pickup is larger than $th$. In this case, we can extract the signal based on the following equation.

$$X'(\tau,\omega) = \begin{cases} X(\tau,\omega) & Y(\tau,\omega) > th \\ 0 & Y(\tau,\omega) < th \end{cases} \qquad \text{(Eq. 10)}$$
$$= M_1(\omega) S(\tau,\omega)$$

When we play the concert with many instruments, the harmonic overtones of the instrumental sound often overlap, that is, Eq. 5 is not satisfied. In such case, the output of the proposal method is altered as follows and we can partially reduce the noise:

$$X'(\tau,\omega) = \begin{cases} M_1(\omega) S(\tau,\omega) + \sum_i M_{2i}(\omega) N_i(\tau,\omega) & \text{if } Y(\tau,\omega) > th \\ 0 & \text{if } Y(\tau,\omega) < th \end{cases} \qquad \text{(Eq. 11)}$$

### 2.2 Temporally smooth noise reduction

The processing method given in Eq.11 performs the frequency component switching according to the value of $Y(\tau, \omega)$. This switching causes a noise on each frequency component. Therefore, the switching should be gradual. To realize the smooth switching, we introduce a reduing coefficient $a(\tau, \omega)$ as follows:

$$X'(\tau,\omega) = a(\tau,\omega) \cdot X(\tau,\omega)$$
$$a(\tau,\omega) = \begin{cases} d \cdot a(\tau-1,\omega) & Y(\tau,\omega) < th \\ 1 - r + r \cdot a(\tau-1,\omega) & Y(\tau,\omega) < th \end{cases} \qquad \text{(Eq.12)}$$
$$a(0,\omega) = 1$$
$$\text{where} \quad 0 < d < 1, \quad 0 < r < 1$$

It means that $a(\tau, \omega)$ is increased or decreased exponentially as shown in Figs. 2 and 3. $d$ and $r$ determine the time constants of the decay and rise of $a(\tau, \omega)$, respectively. The smaller $d$ is, the more rapidly the noise is reduced. The bigger the $r$ is, the more rapidly the signal becomes to be passed. We determine the parameter $d$ and $r$ empirically to get the high quality impression sound.
The output of proposal method $x'(t)$ is obtained by IFFT of $X'(t)$ described as follows;

$$x'(t) = \mathcal{F}^{-1}[X'(\tau,\omega)] = \mathcal{F}^{-1}[\, a(\tau,\omega) \cdot X(\tau,\omega)] \equiv G[x(t)] \qquad \text{(Eq. 13)}$$

where $\mathcal{F}^{-1}$ represents IFFT. $G[\bullet]$ represents the function of the proposal method.

Table 1.- The parameter of FFT

| Window size | 2,048 |
|---|---|
| Overlap | 75% |
| Window | Hamming |
| Sampling frequency | 44.1kHz |

Table 2.- The SNR in simulation experiment

|  | SNR [dB] |
|---|---|
| $SNR_o$ | -2.55 |
| $SNR_p$ | 6.15 |

Table 3.- The SDR in simulation experiment

|  | SDR [dB] |
|---|---|
| $SDR_m$ | 16.34 |
| $SDR_p$ | 0.44 |

## 3. EXPERIMENTS
### 3.1 SIMULATION EXPERIMENT

To evaluate the proposal method quantitatively, we performed two simulated experiments; the experiments to estimate the Signal to Noise Ratio (*SNR*) and the Signal Distortion Ratio (*SDR*). We prepared the data simulating the signals recorded by the microphone and the piezoelectric pickup. In what follows, we call these data as the simulated microphone data and the simulated piezoelectric pickup data. We utilize the KOTO sound as a signal and utilize the other instruments sounds including the electric guitar, the electric bass, and drums as a noise. We obtain the microphone data by mixing the signal and the noise at the similar level and applying reverb control to simulate the echoic environment. We obtain the piezoelectric pickup data by the signal and quite law level the noise and applying equalizing control to simulate the piezoelectric pickup distortion. Table 1 shows the parameters of FFT.

We performed simulated experiment to estimate the SNR. Let us define $SNR_o$ and $SNR_p$ as the SNR before the process and after the process respectively. $SNR_o$ and $SNR_p$ can be described as follows:

$$SNR_O = 10\log_{10} \frac{\sum_{t=0}^{T} s(t)^2}{\sum_{t=0}^{T} n(t)^2} \quad \text{(Eq.14)}$$

$$SNR_P = 10\log_{10} \frac{\sum_{t=0}^{T} G[s(t)]^2}{\sum_{t=0}^{T} G[n(t)]^2} \quad \text{(Eq.15)}$$

where $s(t)$ and $n(t)$ represent the signal and the noise of the simulated microphone data, respectively. $T$ represents the data length. $G[s(t)]$ and $G[n(t)]$ represents the outputs of the proposal method in case that we utilize only the signal and the noise, respectively. As shown in Table 2, the *SNR* is improved by 8.7 dB. Let us define $SDR_m$ as the *SDR* after the process. $SDR_m$ is described as follows;

$$SDR_m = 10\log_{10} \frac{\sum_{t=0}^{T} s(t)^2}{\sum_{t=0}^{T} (s(t) - G[s(t)])^2} \quad \text{(Eq.16)}$$

For reference, we also define $SDR_p$ as the *SDR* of the signal of the simulated piezoelectric pickup data. $SDR_p$ is described as follows;

$$SDR_p = 10\log_{10} \frac{\sum_{t=0}^{T} s(t)^2}{\sum_{t=0}^{T} (s(t) - p(t))^2} \quad \text{(Eq.17)}$$

where $p(t)$ represents the signal of the simulated piezoelectric pickup data in the discrete time *t*. We can not compare these values easily because $p(t)$ is not actually calculated from $s(t)$. In this paper, we decided $p(t)$ so that the loudness of $s(t)$ and $p(t)$ become the same level acoustically. Table 3 shows the results.

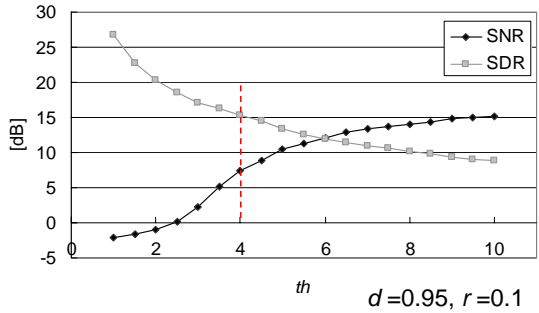
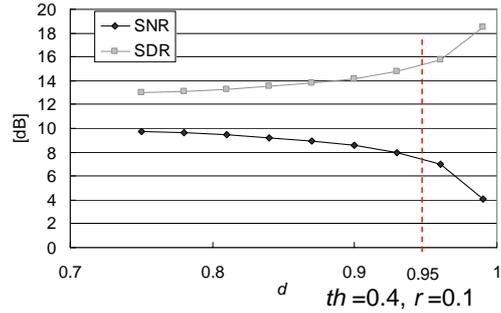

Figure 4.-The effect of *th* on *SNR* and *SDR*    Figure 5.-The effect of *d* on *SNR* and *SDR*

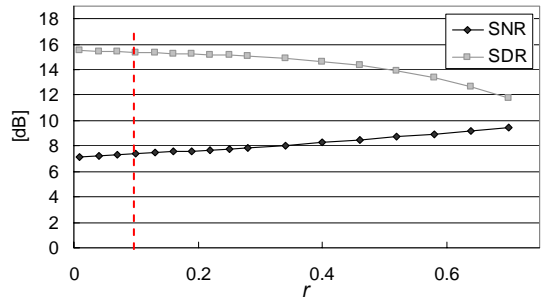

Figure 6.-The effect of *r* on *SNR* and *SDR*

Table 4.- The *SNR* in semi actual environment

|  | SNR [dB] |
|---|---|
| $SNR_o$ | 1.04 |
| $SNR_p$ | 5.76 |

Table 5.- The *SDR* in semi actual environment

|  | SDR [dB] |
|---|---|
| $SDR_m$ | 9.70 |
| $SDR_p$ | -1.05 |

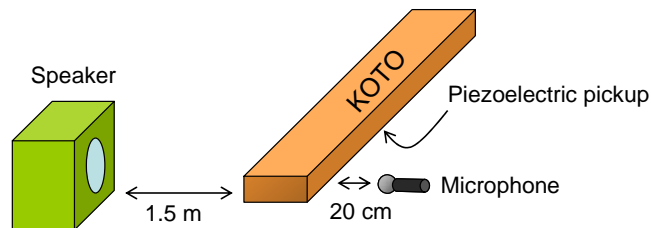

Figure 7. The experimental environment

### 3.2 CONSIDERATION OF PARAMETER
In order to quantitatively estimate the effect of parameter *th*, *d* and *r*, We estimate the relation of these values and the *SNR* and *SDR*. Figs. 4,.5 and 6 show the effect of *th*, *d* and *r* on the *SNR* and *SDR* respectively. These figures shows that there are the trade-off relationship between the *SNR* and the *SDR*. We decided the parameter to obtain the high-quality aurally as follow; *th* is 4, *d* is 0.95, *r* is 0.1. The high-quality impression means that we are not so concerned about the distortion of the output sound and feel that the noise is reduced sufficiently.

### 3.3 SEMI ACTUAL ENVIROMENT EXPERIMENT
We estimate the SNR and SDR of the semi actual data. The semi actual data are prepared by mixing the signal recorded by the actual microphone and the piezoelectric pickup and the noise recorded in different time. The signal is the KOTO sound. The noise contains the drums, the electric bass, and the electric guitar. The parameter *th, d* and *r* were determined manually. Table 4 and 5 show the *SNR* and the *SDR*. The *SNR* was improved by 4.72 dB while the *SDR* is more than 9.7 dB.

### 3.4 ACTUAL ENVIROMENT EXPERIMENT
We also processed the proposal method with the actual environment data. The data was recorded by the microphone at the experimental environment as shown in Fig. 7. When we play the KOTO, the sound pressure at the position of the microphone is 83~93 dB. When we play the background sound with the speaker, the sound pressure at the same position is 85~90 dB. The sound pressure is measured by the sound level meter.

Although we can not estimate the *SNR* and *SDR*, we could obtain the same level *SNR* as simulation experiment acoustically.

## 4. CONCLUSIONS

In this paper, we paid attention to the characteristic of piezoelectric device and proposed the noise reduction combining microphone and piezoelectric device. The merits of the proposal method are summarized as follows;

(1) We do not need multiple microphones and do not need to know the position of microphone. We only have to set one microphone and one piezoelectric pickup at anywhere we would like to set.

(2) We can utilize the proposal method even if we can not the binary mask approach because the noise is so laud.

(3) We can utilize the proposal method even if the signal and the noise come from the same direction.

The proposal method improved the SNR by 8.7 dB and the SDR of this method is 13.64 dB in the simulation experiment.

In the future, we will realize the proposal method in real-time system as well as the parameter optimization. We also aim at developing the time domain process such as the band pass filter instead of the reducing process in the frequency domain.

## 5. ACKNOWLEDGMENT

This research is supported by CREST project "Foundation of technology supporting the creation of digital media contents" of JST. This research was also supported in part by "Establishment of Consolidated Research Institute for Advanced Science and Medical Care", Encouraging Development Strategic Research Centers Program, the Special Coordination Funds for Promoting Science and Technology, Ministry of Education, Culture, Sports, Science and Technology, Japan. This research was also supported (in part) by the Grant-in-Aid for the WABOT-HOUSE Project by Gifu Prefecture and the 21st Century Center of Excellence Program, "The innovative research on symbiosis technologies for human and robots in the elderly dominated society", Waseda University.

**References**:
[1] K. Sasaki and K Hirata, "3D-localization of a stationary random acoustic source in near-field by using 3 point-detectors," Trans of SICE, Vol.34, No.10, pp.1329-1337, 1998.
[2] Y. Yamasaki and T. Itow, "Measurement of spatial information in sound fields by the closely located four point microphone method," J. Acoustic Society Japan, Vol.10, No.2, pp.101-110, 1990.
[3] M. Matsumoto and S. Hashimoto, "A miniaturized adaptive microphone array under directional constraint utilizing aggregated microphones," J. Acoustic Society America, Vol.119, No.1, pp.352-359, 2006.
[4] K. Kiyohara, Y. Kaneda, S. Takahashi, H. Nomura, and J. Kojima, "A microphone array system for speech recognition," Proc. IEEE Int'l Conf. on Acoustics, Speech, and Signal Processing, pp.215-218, 1997.
[5] Y. Kaneda and J. Ohga, "Adaptive microphone array system for noise reduction," IEEE Trans on Acoust. Speech Source Process, Vol.ASSP-34, No.6, pp.1391-1400, 1986.
[6] K. Takao, M. Fujita, and T. Nishi, "An adaptive antenna array under directional constraint," IEEE Trans on Antennas Propagat., Vol.24, pp.662-669, 1976.
[7] A. J. Bell and T. J. Sejowski, "An information maximization approach to blind separation and blind deconvolution," Neural Comput, vol.7, pp.1129-1159, 1995.
[8] H. Saruwatari, S. Kurita and K. Takeda, "Blind source separation combining frequency-domain ICA and beamforming," Proc. Int'l Conf. on Acoustics, Speech, and Signal Processing, pp.146-157, 2001.
[9] T. Ihara, M. Handa, T. Nagai and A. Kurematsu, "Multi-channel speech separation and localization by frequency assignment" IEICE trans on Fundamentals, vol.J86-A, No.10, pp.998-1009, 2003.
[10] S. Rickard and O. Yilmaz, "On the approximate w-disjoint orthogonality of speech," Proc. Int'l Conf. on Acoustics, Speech, and Signal Processing, pp.529-532, 2002.
[11] M. Aoki, Y. Yamaguchi, K. Furuya, and A. Kataoka, "Modifying SAFIA:Separation of the Target Source Close to the Microphones and Noise Sources Far from the Microphones," IEICE trans on Fundamentals, Vol.J88-A, No.4, pp.468-479, 2005.